\documentstyle[12pt,epsf]{article}

\begin{document}

\baselineskip=18pt plus 0.2pt minus 0.1pt

\makeatletter

\def\@bgnmark{<}
\def\@endmark{>}
\def\WKht{.85}
\def\WKsep{.4}
\def\WKrule{.03}
\newcount\@bgncnt
\newcount\@endcnt
\newcount\@h@ight
\newcount\TempCount
\newif\if@Exist
\newdimen\@tempdimc
\newdimen\@tempdimd
\newdimen\h@ight 
\newdimen\w@dth  
\def\sqrt{\radical"270370}

\def\SEPbgn#1<#2#3#4\@@{\xdef\@MAE{#1}\xdef\@MARK{#2}
\xdef\@FRONT{#3}\xdef\@USIRO{#4}}
\def\SEPend#1>#2#3#4\@@{\xdef\@MAE{#1}\xdef\@MARK{#2}
\xdef\@FRONT{#3}\xdef\@USIRO{#4}}
\def\c@lc{
 \setbox0=\hbox{$\displaystyle \@FRONT$}
 \@tempdima\wd0 \@tempdimb\ht0
 \settowidth{\@tempdimc}{$\displaystyle \@MAE$}
 \settowidth{\@tempdimd}{$\displaystyle \@list$}
 \divide\@tempdima by2 \advance\@tempdima by \@tempdimc
 \advance\@tempdima by \@tempdimd}

\def\@dblfornoop#1\@@#2#3#4{}
\def\@dblfor#1;#2:=#3\do#4{\xdef\@fortmp{#3}\ifx\@fortmp\@empty \else%
 \expandafter\@dblforloop#3\@nil,\@nil,\@nil\@@#1#2{#4}\fi}
\def\@dblforloop#1,#2,#3\@@#4#5#6{\def#4{#1} \def#5{#2}%
 \ifx #4\@nnil \let\@nextwhile=\@dblfornoop \else%
 #6\relax \let\@nextwhile=\@dblforloop\fi\@nextwhile#3\@@#4#5{#6}}

\def\fin@endpt#1#2{
\@dblfor\MemBer;\NextmemBer:=#2\do{\def\@bject{#1}%
 \if \MemBer\@bject \xdef\@endpt{\NextmemBer} \@Existtrue\fi}}%
\def\fin@h@ight#1#2{
 \@tempcnta\z@%
 \@tfor\MEmber:=#2\do{\advance\@tempcnta\@ne%
 \ifnum \@tempcnta=#1 \@h@ight=\MEmber\fi}}

\def\wicksymbol#1#2#3#4#5{
 \@tempdima=#3 \advance\@tempdima-#1%
 \@tempdimc=#5\h@ight \@tempdimb=\@tempdimc \advance\@tempdimb-\w@dth%
 \@tempdimd=#2 \advance\@tempdimd-1.587ex
 \hskip#1%
 \vrule height \@tempdimc width\w@dth depth-\@tempdimd \kern-\w@dth%
 \vrule height \@tempdimc width\@tempdima depth-\@tempdimb\kern-\w@dth%
 \@tempdimd=#4 \advance\@tempdimd-1.587ex 
 \vrule height \@tempdimc width\w@dth depth-\@tempdimd}

\def\first#1{\expandafter\@mae#1\@nil}
\def\secnd#1{\expandafter\@ato#1\@nil}
\def\@mae#1;#2\@nil{#1}
\def\@ato#1;#2\@nil{#2}

\def\wick#1#2{%
 \h@ight=\WKht ex \w@dth=\WKrule em%
 \def\@wickdata{} \def\bgnend@list{} \@bgncnt\z@ \@endcnt\z@%
 \def\@list{} \def\bgnp@sition{} \def\endp@sition{}%
 \xdef\str@ng{#2}
 \@tfor\m@mber:=#2\do{%
 \ifx\m@mber\@bgnmark \advance\@bgncnt\@ne
  \expandafter\SEPbgn\str@ng\empty\@@ \c@lc 
  \xdef\bgnp@sition{\bgnp@sition\@MARK,\the\@tempdima;\the\@tempdimb,}
  \xdef\@list{\@list\@MAE\@FRONT}
  \xdef\str@ng{\@USIRO}\fi
 \ifx \m@mber\@endmark \advance\@endcnt\@ne
  \expandafter\SEPend\str@ng\empty\@@ \c@lc
  \xdef\endp@sition{\endp@sition\@MARK,\the\@tempdima;\the\@tempdimb,}
  \xdef\@list{\@list\@MAE\@FRONT}
  \xdef\str@ng{\@USIRO}\fi}
  \xdef\@list{\@list\@USIRO}
 \ifnum\@bgncnt=\@endcnt \else%
 \@latexerr{The numbers of `<' and `>' do not match}%
 {You have written different numbers of < and >}\fi%
 \TempCount\z@ \@tfor\mmbr:=#1\do{\advance\TempCount\@ne}%
 \ifnum\@bgncnt=\TempCount \else%
 \@latexerr{The number of numbers in the first argument is different
 with that of contractions <...>}%
 {Give the same numbers of heights as the contractions <...>}\fi
 \mathop{\vbox{\m@th\ialign{##\crcr\noalign{\kern\WKsep ex}%
 $\m@th \TempCount\z@%
 \@dblfor\member;\nextmember:=\bgnp@sition\do{
 \advance\TempCount\@ne \xdef\@bgnpt{\nextmember}%
 \@Existfalse%
 \fin@endpt{\member}{\endp@sition}%
 \if@Exist \else \@latexerr{The begin-mark `<\member' has no
corresponding end-mark `>\member'}{You should write coinciding label 
like <\member .. >\member}\fi%
 \fin@h@ight{\TempCount}{#1}%
 \setbox0=\hbox{%
 $\wicksymbol{\first\@bgnpt}{\secnd\@bgnpt}{\first\@endpt}%
 {\secnd\@endpt}{\@h@ight}$\hss}
 \dp0\z@ \wd0\z@ \box0%
 }$\crcr\noalign{\kern\WKsep ex\nointerlineskip}%
 \setbox0=\hbox{$\displaystyle\@list$}\ht0=1.587ex%
 \box0\crcr}}}\limits}

\@addtoreset{equation}{section}
\renewcommand{\theequation}{\thesection.\arabic{equation}}

\newcommand{\nn}{\nonumber}
\newcommand{\bF}{\ket{B(F)}}
\newcommand{\BF}{B(F)}
\newcommand{\bFx}{\ket{B\Bigl(F(x)\Bigr)}}
\newcommand{\BFx}{B\Bigl(F(x)\Bigr)}
\newcommand{\IF}{I(F)}
\newcommand{\IFx}{I_{\rm new}\Bigl(F(x)\Bigr)}
\newcommand{\bnew}{\ket{{\cal B}}}
\newcommand{\Bnew}{{\cal B}}
\newcommand{\bnewA}{\ket{{\cal B}\Bigl(A_\mu(x)\Bigr)}}
\newcommand{\BnewA}{{\cal B}\Bigl(A_\mu(x)\Bigr)}
\newcommand{\Lz}{\Lambda\Bigl(\zeta_\mu(x)\Bigr)}
\renewcommand{\star}{*}
\newcommand{\tr}{\mathop{\rm tr}}
\newcommand{\QB}{Q_{\rm B}}
\newcommand{\Half}{\frac{1}{2}}
\newcommand{\bra}[1]{\left\langle #1\right|}
\newcommand{\hs}[1]{\hspace*{#1}}
\newcommand{\vs}[1]{\vspace*{#1}}
\newcommand{\ket}[1]{\left| #1\right\rangle}
\newcommand{\VEV}[1]{\left\langle #1\right\rangle}
\newcommand{\braket}[2]{\VEV{#1 | #2}}
\newcommand{\ac}{\overline{c}}
\newcommand{\calO}{{\cal O}}
\newcommand{\p}{\partial}
\newcommand{\wt}[1]{\widetilde{#1}}
\newcommand{\wtX}{\wt{X}}

\makeatother

\begin{titlepage}
\title{
\hfill\parbox{4cm}
{\normalsize KUNS-1598\\{\tt hep-th/9909027}}\\
\vspace{1cm}
Corrections to D-brane Action\\and Generalized Boundary State
\\[50pt]
}
\author{
{\sc Koji Hashimoto}\thanks{{\tt hasshan@gauge.scphys.kyoto-u.ac.jp}}
\\[7pt]
{\it Department of Physics, Kyoto University, Kyoto 606-8502, Japan} 
}
\date{\normalsize September, 1999}
\maketitle
\thispagestyle{empty}

\begin{abstract}
\normalsize

In this paper, we generalize a boundary state to the one incorporating
non-constant gauge field strength as an external background coupled to
the boundary of a string worldsheet in bosonic string theory. This
newly defined boundary state 
satisfies generalized nonlinear boundary conditions with non-constant
gauge field strength, and is BRST invariant. The divergence immanent
in this boundary state coincide with the one calculated in a string
$\sigma$ model. We extract the relevant massless part of this
generalized boundary state, and give a part of the D-brane action with
the non-constant gauge field strength, that is, derivative corrections
to the D-brane action.

\end{abstract}

\end{titlepage}

\section{Introduction}
\label{sec:intro}

D-branes \cite{DLP} are indispensable objects in string theories to 
the understanding of their non-perturbative aspects, including various
dualities \cite{HT}. One way to examine string dualities  is to use
the D-brane action which is a low energy effective action of massless
fields induced on the D-brane \cite{Leigh}. For example, taking the
worldvolume dual of the  D$1$-brane action, we obtain a Nambu-Goto
string action with a tension rotated by $SL(2,Z)$ duality
transformation which is conjectured to  exist in type IIB superstring
theory \cite{Sch}. Similar analyses have  been performed for other
cases, such as D$p$-brane ($2\!\le\!p\!\le\! 4$) actions or
super-$p$-brane actions \cite{Sch,others}.  As another argument on
string dualities using D-brane action we cite ref.\ \cite{Matsuo},
which shows that the BPS mass spectra given as a volume factor  of the
D-brane action describing compactified D-branes are consistent with
U-duality.

The D-brane actions used in the above analyses are the ones given by
the lowest order calculation in the string $\sigma$ model approach
\cite{dorn,ACNY,CLNY,Ber}, or in the open string partition function
approach \cite{Par}, in addition to the ordinary scattering amplitude
approach \cite{scatt}. Higher order calculations of the D-brane
actions have been known through the latter approaches \cite{Tse}, in
the trivial background of the massless sector of the closed string
excitations (flat metric and no Kalb-Ramond two-form field).  
This adds to the D-brane action higher derivative 
($\alpha'$) corrections,\footnote{
There are some arguments that the supersymmetry ensures the absence
of correction terms to the D-brane action \cite{Bagger}.
} 
which, however, may ruin the dualization procedure
of refs.\ \cite{Sch,others}.\footnote{ 
T-duality for the bulk supergravity action can be extended to the one
including the higher loop effects (see ref.\ \cite{cordual}). In other
references, the higher derivative corrections to the D-brane action in
the gravity sector (especially, (curvature)$^2$ terms) are calculated
and its consistency with string dualities is checked \cite{Bach}. 
}
Therefore, a systematic framework which gives higher derivative
corrections to the D-brane action is highly desired.

Since D-branes are defined as hypersurfaces on which the boundaries of
the string worldsheets can sit, the boundary states
\cite{bouf,CLNY2,CLNY3,PC,Ishi} 
describe the dynamics of the D-branes well. A part of the D-brane
action (more precisely, couplings to the fields of the closed string
excitations) is encoded in the boundary state \cite{BD}. However, this
boundary state is defined as an eigen state of the boundary conditions
for the oscillating modes of the attached string, the explicit form of
the boundary state has been given only for simple backgrounds which 
give linear boundary conditions. One of those configurations is the
constant field strength on the D-brane. Therefore, boundary states
including general configuration 
of the boundary gauge field give us a method to calculate the higher 
derivative corrections to the D-brane action.

In ref.\ \cite{our}, a D-brane sector was introduced to the system of
covariant closed bosonic string field theory (SFT) \cite{HIKKO}. 
The D-brane action corresponds to the source term in this SFT, and the
boundary state is a constituent of it. 
The dynamical degree of freedom of the boundary state adopted there
was only the constant field strength.
In this scheme of SFT, a $\sigma$ model gauge transformation
was realized is such a way that it yields a constant shift of the
gauge field strength in the boundary state.

In this paper, we present a generalized boundary state, which is
constructed by a gauge transformation in the SFT.\footnote{We
  concentrate on the bosonic string theory in this paper. A work on
  supersymmetric extension will appear soon \cite{KH}.} 
This boundary state incorporates the non-constant modes of
$F_{\mu\nu}(x)$ and gives the derivative corrections to the D-brane
actions.  
Our new boundary state is defined for an arbitrary gauge field
configuration $A_\mu(x)$, and we carry out the explicit calculations
up to the second order in the derivatives on $F_{\mu\nu}(x)$. 
This boundary state includes divergences, and the requirement for
these divergences to vanish gives a constraint on the configuration of
the gauge field $A_\mu(x)$. This constraint is shown to be identical
with the vanishing of the $\beta$ function in a string $\sigma$ model, 
hence we regard this constraint as an equation of motion for the gauge
field. After this regularization procedure for the boundary state, we
obtain the higher derivative corrections to the D-brane action, by
extracting relevant parts in the boundary state. It is expected 
that this generalized boundary state serves as an useful tool for the
investigation of the dualities or other dynamics in string theory. 

The organization of the rest of this paper is as follows. In the next
section, after summarizing the properties of the boundary states
adopted in ref.\ \cite{our}, this boundary state is generalized to the
one which has non-constant $F_{\mu\nu}(x)$ boundary coupling. 
In sec.\ \ref{sec:3}, we study the divergence immanent in the
generalized boundary state, and see the coincidence with a string
$\sigma$ model. 
In sec.\ \ref{sec:correction}, using the new boundary state proposed
in  sec.\ \ref{sec:extending}, we calculate the correction to the
D-brane action.
The final section is devoted to our conclusion and discussions. 
We include an appendix where the calculation of the formula appearing
in this paper is given. 


\section{Generalizing the boundary state}
\label{sec:extending}

As mentioned in the introduction, a certain sector of the boundary
state is related to the low energy effective action of D-branes
\cite{BD}. 
The boundary state adopted in ref.\ \cite{our} and in most of
the literatures has a single free parameter, that is, {\it constant}
field strength of the gauge field on the D-brane.\footnote{D-branes
  moving with a 
  constant velocity or tilted by a constant angle are treated as a
  T-dual (or boosted) version of the case with the constant field
  strength \cite{front,move}.}
 Therefore the obtained effective action
on the D-brane 
contains only the constant field strength. In this section, we
generalize the boundary state so that it includes {\it non-constant}
modes of the field strength, in order to calculate the corrections to
the D-brane action. These corrections correspond to the derivative
corrections, since we expand the field strength by derivatives. 

First let us summarize the relevant properties of the boundary state
with the constant field strength. The boundary state $\bF$ is defined
as an eigen state of the boundary conditions for open bosonic strings
\cite{bouf,CLNY2,PC}:
\begin{eqnarray}
&&X^i(\sigma)\bF=0 ,
\label{X_B=0}\\
&&\left(\pi P_\mu(\sigma)+F_{\mu\nu}\p_\sigma
X^\nu(\sigma)\right)\bF=0 ,
\label{(P-FX)B}\\
&&\pi_c(\sigma)\bF=\pi_{\ac}(\sigma)\bF=0 .
\label{pi_c_B}
\end{eqnarray}
The BRST invariance of the boundary state is a consequence of eqs.\
(\ref{X_B=0}) --- (\ref{pi_c_B}).
The oscillator representation of $\ket{B(F)}$ reads
\begin{eqnarray}
\ket{\BF(x^M,\bar{c}_0,\wt{\alpha})}= 
N(F)\ket{B_{\rm N}(F)}\otimes\ket{B_{\rm D}}\otimes\ket{B_{\rm gh}},
\label{bou}
\end{eqnarray}
where
\begin{eqnarray}
&&\ket{B_{\rm N}(F)}=\exp \biggl\{ 
-\sum_{n\geq 1} {1\over n}
\alpha_{-n}^{(-)\mu}\calO (F)_\mu^{\;\; \nu}\alpha_{-n\;\nu}^{(+)}
\biggr\}\ket{0}_{p+1},\\
&&\ket{B_{\rm D}}=\exp \biggl\{ 
\sum_{n\geq 1} {1\over n}
\alpha_{-n}^{(-)i}\alpha_{-n\; i}^{(+)}
\biggr\}\ket{0}_{d-p-1}\delta^{d-p-1}(x^i),\\
&&\ket{B_{\rm gh}}=\exp \biggl\{ 
\sum_{n\geq 1} 
(c_{-n}^{(-)}\bar{c}_{-n}^{(+)}+c_{-n}^{(+)}\bar{c}_{-n}^{(-)})
\biggr\}\ket{0}_{\rm gh}.\label{ghostF}
\end{eqnarray}
The orthogonal matrix $\calO$ is defined as $\calO_\mu^{\;\;\nu}=
(1-F)_\mu^{\;\;\rho}\{(1+F)^{-1}\}_{\rho}^{\;\;\nu}$.
Although the front factor $N(F)$ is arbitrary, this can be fixed by
various ways \cite{BD,our,front}. Here we shall follow ref.\
\cite{our}, where the gauge invariance principle for the D-brane
(source) term in the closed SFT action has fixed the front
factor. Only when  the front factor is given by
\begin{eqnarray}
N(F)=-{T_p\over 4}\Bigl(\det (1+F)\Bigr)^{-\zeta(0)},
\label{nf}
\end{eqnarray}
the boundary state is subject to the following SFT gauge
transformation:
\begin{eqnarray}
\label{b*f}
\ket{\Lambda\star \BF} = \delta_\Lambda \ket{\BF},
\end{eqnarray}
where the $\star$ product denotes three string interaction vertex.
A string field $\Lambda$ is a gauge transformation parameter
\begin{eqnarray}
\ket{\Lz}=-i\bar{c_0}\zeta_\mu (x)\left(
\bar{c}_{-1}^{(+)}\alpha_{-1}^{(-)\mu}
-\bar{c}_{-1}^{(-)}\alpha_{-1}^{(+)\mu}
\right)\ket{0}
\label{Lambda}
\end{eqnarray}
with
\begin{eqnarray}
\label{zetax}
\zeta_\mu (x)=\zeta_{\mu\nu}x^\nu,
\end{eqnarray}
and we have defined the gauge transformation for the variable $F$ as 
\begin{eqnarray}
\delta_\Lambda F_{\mu\nu} = \zeta_{\mu\nu} - \zeta_{\nu\mu}.  
\end{eqnarray}
For verifying the gauge transformation (\ref{b*f}), we have used a
relation on the $\star$ product
\begin{eqnarray}
\ket{\Lambda\star \BF}
=-{i\over 2\pi}\oint\! d\sigma 
\, \p_\sigma X^\mu \zeta_{\mu\nu} X^\nu \bF ,
\label{precon}
\end{eqnarray}
for the above $\Lambda$ (\ref{Lambda}). 

The gauge transformation for the closed string field is given by 
\begin{eqnarray}
\delta_\Lambda\Phi = Q_{\rm B}\Lambda +2\Phi\star\Lambda,   
\label{transP}
\end{eqnarray}
therefore with the transformation parameter (\ref{Lambda}), the
inhomogeneous part of the gauge transformation (\ref{transP})
generates a shift of a component Kalb-Ramond gauge field
$b_{\mu\nu}(x)$ in $\Phi$ by $\p_\mu\zeta_\nu (x) -\p_\nu\zeta_\mu
(x)$. (Our notation of  
SFT is given in ref.\ \cite{our}.) This fact is expected from the
gauge symmetry of the string $\sigma$ model, which is defined as
$\delta_{(\sigma {\rm model})} A_\mu(x) = -\zeta_\mu(x)$ and
$\delta_{(\sigma {\rm model})} 
b_{\mu\nu}(x) = \p_\mu\zeta_\nu(x) - \p_\nu\zeta_\mu (x)$\cite{CLNY}.
Note that in the string $\sigma$ model approach, this gauge
transformation $\delta_{(\sigma {\rm model})}$ generates all the modes
of the gauge field $A_\mu(x)$ on the boundary. Therefore, in order to
generalize the boundary state to the one incorporating the functional
degree of freedom $A_\mu(x)$ (or $F_{\mu\nu}(x)$), it is natural to
demand the following relation for the new boundary state
$\ket{\BnewA}$ : 
\begin{eqnarray}
\label{transnew}
\delta_\Lambda \bnew
=-2 \ket{\Lambda \star \Bnew}.
\end{eqnarray}
In this relation the parameter string field $\Lambda$ is defined by
(\ref{Lambda}) for an {\it arbitrary} function $\zeta_\mu(x)$, and 
the gauge transformation law concerning $A_\mu(x)$ is 
\begin{eqnarray}
\label{gaugetr}
\delta_\Lambda A_\mu (x) = - \zeta_\mu(x).
\end{eqnarray}

Since the gauge transformation (\ref{gaugetr}) generates an arbitrary
deformation of $A_\mu(x)$ by the parameter $\zeta_\mu(x)$, $\bnewA$ is
obtained by ``integrating'' eq.\ (\ref{transnew}). For this purpose we
use the following formula for the star product $\Lz\star \Psi$ between
the present $\Lambda$ and an arbitrary string functional $\Psi$ (the 
derivation is given in app.\ \ref{app:star}):
\begin{eqnarray}
\label{star}
-2\ket{\Lz\star \Psi}
=\left\{
{i\over\pi}\int_0^{2\pi}\!\!d\sigma\;\p_\sigma X^\mu
\zeta_\mu(X) +\int_0^{2\pi}\!\!d\sigma\;\p^\mu \zeta_\mu(X) \;
i\pi_c{\Big |}_{\rm oscil.} \; i\pi_{\bar c}{\Big |}_{\rm oscil.}
\right\}\ket{\Psi}.
\end{eqnarray}
This is a generalization of the identity (\ref{precon}). Assuming that
the ghost part of the new boundary state is given by the same form
(eq.\ (\ref{ghostF})) as before, the second term on the RHS of
eq.\ (\ref{star}) acting on the state $\bnew$ vanishes due to the
ghost boundary condition (\ref{pi_c_B}). Then the desired generalized
boundary state $\bnew$ is obtained by exponentiating the 
operator on the RHS of (\ref{star}):\footnote{\label{weight}
  This definition is very natural in the
  sense of path-integral approach of string $\sigma$ model
  \cite{CLNY2}. The boundary state is regarded as an initial or final
  wave functional for the path-integral. The rotational operator
  $U[A]$ is a path-integral weight $\exp (iS^{(\sigma)}_{\rm int})$,
  where $S^{(\sigma)}_{\rm int}$ is a minimal interaction term with
  external gauge fields $A_\mu (X)$ in the string $\sigma$ model
  action. }
\begin{eqnarray}
\label{Bnew}
\bnewA\equiv U[A]\:\ket{B(F\!=\!0)},
\end{eqnarray}
with
\begin{eqnarray}
\label{U[A]}
  U[A]\equiv \exp \left({-i\over \pi}\oint\! d\sigma 
                  \, \p_\sigma X^\mu A_{\mu}(X)
  \right).
\end{eqnarray}
Note that $U[A]$ and hence $\Bnew$ has an invariance under the $U(1)$
gauge transformation of $A_\mu(x)$, $A_\mu\to A_\mu +\p_\mu\epsilon$.

The new boundary state (\ref{Bnew}) is reduced to the previous one
(\ref{bou}) when the dynamical variable is restricted to the constant
field strength
\begin{eqnarray}
  \label{Aconst}
  A_\mu (x) \equiv a^{(1)}_{\mu\nu}x^\nu\;\;,\;\;\;
  -\left(a^{(1)}_{\mu\nu}-a^{(1)}_{\nu\mu}\right)=F_{\mu\nu},
\end{eqnarray}
since the Neumann part of the previous boundary state (\ref{bou}) is
written in a form 
\begin{eqnarray}
N(F)\ket{B_{\rm N}(F)}&=&
  \exp \left({i\over 2\pi}\oint\! d\sigma 
                  \, \p_\sigma X^\mu F_{\mu\nu} X^\nu 
  \right)\ket{B_{\rm N}(F\!=\!0)}\label{divideF}\\
&=&
  \exp \left({-i\over \pi}\oint\! d\sigma 
                  \, \p_\sigma X^\mu A_{\mu}(X)
  \right)\ket{B_{\rm N}(F\!=\!0)}.\label{rewrite}
\end{eqnarray}

Our $\bnew$ satisfies various desired properties. One of them is
the fact that it obeys the following generalized {\it nonlinear}
boundary condition 
\begin{eqnarray}
\label{eigen}
\biggl[
\pi P_\mu + F_{\mu\nu}(X)
\p_\sigma X^\nu \biggr] \bnew = 0,
\end{eqnarray}
where the field strength $F_{\mu\nu}(X)$ is defined as usual:
\begin{eqnarray}
\label{generalF}
F_{\mu\nu}(X)=\p_\mu A_\nu(X) -\p_\nu A_\mu(X).
\end{eqnarray}
This eigen equation (\ref{eigen}) is easily proved from a property of 
the unitary operator $U[A]$ (\ref{U[A]}):
\begin{eqnarray}
\label{Uproperty}
U[A]\pi P_\mu U[A]^{-1}
=\pi P_\mu + F_{\mu\nu}(X) \p_\sigma X^\nu
\end{eqnarray}
and $P_\mu \ket{B(F\!=\!0)}=0$.
The new boundary state $\bnew$ satisfies a generalized boundary
condition (\ref{eigen}) which is in the same form required in the
scheme of the string $\sigma$ model \cite{ACNY,CLNY,CLNY2}. Thus also
from this point of view it is
possible to assert that this state $\bnew$ really describes a boundary
of a string worldsheet coupled to a massless gauge field $A_\mu(x)$.

One of the other properties of $\bnew$ is the BRST invariance which is 
important to understand loop-corrected equations of motion for closed
string theory \cite{CLNY2,CLNY3}\footnote{In the closed SFT, the
  D-brane action as well as the usual closed string field action is
  implemented \cite{our} as 
  $    S=\Half \braket{\Phi}{\QB|\Phi} + \braket{\Phi}{B}, $
  where we consider only the linearized part. The equation of motion
  follows as 
  $  \QB\ket{\Phi} + \ket{B}=0,$
  hence multiplying $\QB$ on this expression we obtain a constraint
  $\QB\ket{B}=0$. This constraint of the BRST invariance of the
  boundary state is also deduced from the gauge invariance of the
  whole SFT action\cite{our}.}.
The BRST charge can be written in the form (see Appendix A in the
ref.\ \cite{our} for its complete expression)
\begin{eqnarray}
Q_{\rm B} =2\sqrt{\pi}\oint \! d\sigma
\biggl\{i\pi_{\bar{c}}\cdot \left[
{\rm functions\; of\;}P_M\; {\rm and}\; X^N\right]
-c\; P_M\p_\sigma X^M \biggr\}  \; + \; Q_{\rm B}^{({\rm ghost})},
\end{eqnarray}
where the term $Q_{\rm B}^{({\rm ghost})}$ consists only of the 
ghost coordinates. Noting that the ghost part of the new boundary 
state $\bnew_{\rm ghost}$ is the same as the previous one
$\ket{B_{\rm gh}}$, we have
\begin{eqnarray}
\pi_{\bar{c}}\bnew\; =\; Q_{\rm B}^{({\rm ghost})}
\bnew\;=\; 0. 
\end{eqnarray}
Then the commutativity
\begin{eqnarray}
\label{Acomu}
\biggl[
P_\mu\p_\sigma X^\mu ,
\oint\! d\sigma^\prime \p_{\sigma^\prime} X^\nu A_\nu(X)
\biggr]=0
\end{eqnarray}
ensures the BRST invariance
\begin{eqnarray}
\label{BRST}
Q_{\rm B}\bnew =0.
\end{eqnarray}

Although this boundary state $\bnew$ is a solution of the equation
(\ref{BRST}) for arbitrary configuration of the gauge field, this
solution is formal, in a sense that the boundary state contains
divergence and thus is not well-defined unless regularized. In the
next section we study on this point in detail.

\clearpage


\section{Divergence in the new boundary state}

\label{sec:3}


\subsection{Calculation of the divergence}
\label{subsec:div}

The new boundary state (\ref{Bnew}) for a general (polynomial)
$A_\mu(x)$ is made of products of the operators $X$'s at the same
point $\sigma$ and therefore it inevitably contains divergences.
We shall show in this subsection that these divergences can be
absorbed into the redefinition of the gauge field as 
$A_\mu^{\rm red}(x)\equiv A_\mu(x) + A_{\rm divergent}$. 
We require that the divergent part $A_{\rm divergent}$ to vanish, in
order to obtain a well-defined boundary state. Eventually, constraint
equation is found to be identical with the vanishing of the $\beta$ 
function of the string $\sigma$ model coupling. Thus this requirement of
no-divergence coincides with the condition of the conformal invariance
in the string $\sigma$ model.

For studying the divergences in the new boundary state,
we divide $X^\mu(\sigma)$ into the zero mode
and the non-zero mode part $\wtX^\mu(\sigma)$,
\begin{eqnarray}
  \label{non-zero}
  X^\mu(\sigma) = x^\mu + \wt{X}^\mu(\sigma), 
\end{eqnarray}
and Taylor-expand $A_\mu\left(X(\sigma)\right)$ around the zero mode
$x^\mu$ as
\begin{eqnarray}
A_\mu(X)=A_\mu(x) + \wtX^\nu\;\p_\nu A_\mu(x) +
\frac{1}{2}\wtX^\nu\wtX^\rho\;\p_\nu\p_\rho A_\mu(x)+\cdots .
\label{Aexpand}
\end{eqnarray}
We denote the contribution of the second term on the RHS of
(\ref{Aexpand}) to $U[A]$ (\ref{U[A]}) by $U_0$ and that of the rest
terms by $V$, and express $U[A]$ as
\begin{eqnarray}
U[A]=V U_0.
\end{eqnarray}
Explicitly, we have
\begin{eqnarray}
\label{U_0}
U_0 = \exp\left(\frac{i}{2\pi}
\oint\!d\sigma \;\p_\sigma X^\mu X^\nu F_{\mu\nu}(x)
\right)
\end{eqnarray}
and
\begin{eqnarray}
\label{V}
\lefteqn{V=
  1+{-i\over 3\pi}\oint\! d\sigma 
    \, \p_\sigma \wtX^\mu\wtX^\nu\wtX^\rho\;\p_\rho F_{\nu\mu}(x)
+{-i\over 8\pi}\oint\! d\sigma 
    \, \p_\sigma \wtX^\mu\wtX^\nu\wtX^\rho\wtX^\delta\;
\p_\delta\p_\rho F_{\nu\mu}(x)
}\nn\\
&&\hs{50mm}
   +\Half\left(
{-i\over 3\pi}\oint\! d\sigma 
    \, \p_\sigma \wtX^\mu\wtX^\nu\wtX^\rho\;\p_\rho F_{\nu\mu}(x)
\right)^2
+\cdots
\end{eqnarray}
In (\ref{V}) we have written explicitly only those terms where the
total number of derivatives acting on $F_{\mu\nu}(x)$ is at most two. 

Since the operator $U_0$ acting on $\ket{B(F\!=\!0)}$ generates the
``old'' boundary  
state (\ref{bou}) with $x$-dependent field strength
(cf., eq.\ (\ref{divideF})),
\begin{equation}
U_0\ket{B(F\!=\!0)}=\bFx
\end{equation}
the new boundary state (\ref{Bnew}) is expressed as
\begin{equation}
\bnew = V\bFx.
\label{VB}
\end{equation}
Note that the approximation to the first order with respect to the
derivative acting on the field strength, where $V$ is put to be $1$,
does {\it not} reduce $\bnew$ to the previous (BRST invariant)
boundary state $\bF$ considered in the ref.\ \cite{our}. Actually, the
decomposition of $U[A]$ into $U_0V$ does not respect the BRST
invariance at each order in the expansion of $V$. 

Next, we shall evaluate the representation (\ref{VB}) by expressing it
only by the creation operators acting upon $\bFx$.
Namely, we change all the annihilation operators in $V$ into the
creation ones with use of the Neumann boundary condition
satisfied by $\bFx$:
\begin{eqnarray}
\label{bc}
\left(\alpha^{(-)}_{n\mu} + \calO_\mu^{\;\nu}(x)
\alpha^{(+)}_{-n\;\nu}\right)\bFx=0,
\end{eqnarray}
where $\calO(x)$ is an orthogonal matrix defined by 
\begin{eqnarray}
\calO_\mu^{\;\;\nu}(x) \equiv \left( {1-F(x) \over 1+F(x)}
\right)_\mu^{\;\nu}\;\;,\;\;\calO\calO^T=\calO^T\calO=1.
\end{eqnarray}
The calculation is straightforward and we here present only the
results (see app.\ \ref{app:div} for the detailed calculation). The
unitary operator $V$ reexpressed in terms of creation operators
contains two kinds of divergences,
\begin{eqnarray}
  \label{div}
   \sum_{n=1}^\infty 1 = \zeta(0)\;\; 
,    \;\;  \sum_{n=1}^\infty {1\over n} = \zeta(1).
\end{eqnarray}
Adopting the zeta function regularization, $\zeta(0)=-1/2$,
we find that the $\zeta(1)$ divergence is absorbed into the
redefinition of the field strength $F(x)$:
\begin{eqnarray}
  \label{Bcreate}
  \bnew = \ket{B(F^{\rm red}(x))} 
+ 
\left[
\begin{array}{l}
\mbox{creation operators}\\
\mbox{with finite coefficients}
\end{array}
\right]
\bFx
\end{eqnarray}
where the redefined $F^{\rm red}(x)$ containing the $\zeta(1)$
divergence is given by
\begin{eqnarray}
  \label{Fren}
\lefteqn{F^{\rm red}_{\mu\nu}(x) \equiv F_{\mu\nu}(x)
+ {1\over 2}\zeta(1)\left( {1\over 1\!\!+\!\!F}\right)^{\rho\delta}
\p_\rho\p_\delta F_{\mu\nu}}
\nn\\
 && {}\hs{15mm}+ {1\over 4}\zeta(1)
\left[  
\left\{
\left( 
{1\over 1\!\!+\!\!F}\right)^{\lambda\rho}\!\!\p_\nu F_{\rho\delta}
\left( {1\over 1\!\!+\!\!F}\right)^{\delta\kappa}\!\!
(\p_\lambda F_{\kappa\mu}+\p_\kappa F_{\lambda\mu})
\right\}
-
\left\{ \mu\leftrightarrow\nu\right\}
\right].
\end{eqnarray}
Here we have kept only those terms written explicitly in eq.\
(\ref{V}).\footnote{
The first derivative correction to the boundary state (or the
Born-Infeld action) should contain two derivatives acting on $F$,
since if the corrected action is written by gauge invariant field
strength $F$, the number of indices should be even.}

The redefinition (\ref{Fren}) can be expressed in terms of
gauge field $A_\mu$ as
\begin{eqnarray}
  \label{Aren}
  A^{\rm red}_\mu(x) \equiv A_\mu(x)+\Half \zeta(1) \left(
1\over{1-F(x)^2}\right)^{\lambda\nu}\p_\lambda F_{\nu\mu}(x).
\end{eqnarray}
For the generalized boundary state to be well-defined (not divergent),
the coefficient of the $\zeta(1)$ in eq.\ (\ref{Aren}) should put
equal to zero. Therefore we obtain a constraint on the configuration of
the gauge field
\begin{eqnarray}
\beta(A)\equiv  \left(
1\over{1-F(x)^2}\right)^{\lambda\nu}\p_\lambda F_{\nu\mu}(x)=0.
\label{beta}
\end{eqnarray}
At a glance one recognizes that this $\beta(A)$ is identical with the 
beta function for the gauge field coupling in the string $\sigma$
model \cite{ACNY}. Hence the constraint (\ref{beta}) for the
generalized boundary state coincides with the conformal invariance
condition in the string $\sigma$ model. We shall explain the detailed
correspondence between our formalism and the string $\sigma$ model
approach in the next subsection.


\subsection{Correspondence to the string $\sigma$ model $\beta$ %
  function}  
\label{subsec:sigma}

We have seen in the previous subsection that the divergences
immanent in the newly defined boundary state coincide with 
the result in the string $\sigma$ model. This feature is natural in a
sense that, in the both formalisms, one evaluates the same weight 
$\exp\left(S_{\rm int}^{(\sigma)}\right)$ (see the footnote
\ref{weight}) on the boundary of the string worldsheet. However, two
approaches appear to be clearly different in the following respect:
although the $\sigma$ model approach gives a loop contribution to the 
gauge potential itself, the divergences in the boundary state show up
in the form of the redefinition of the field strength. In this
subsection we study how the divergences in the boundary state can be
understood in the scheme of the string $\sigma$ model. 

Consider a string $\sigma$ model action \cite{dorn,ACNY,CLNY}
\begin{eqnarray}
\label{Ssigma}
S=
{1\over \pi}
  \left[
    \Half \int d^2\sigma \p_\alpha X_M \p^\alpha X^M + \oint d\sigma
    (X^\mu)' A_\mu [X]
  \right],    
\end{eqnarray}
where the prime denotes the derivative with respect to
a worldsheet coordinate $\sigma$. The boundary of the worldsheet is
put at $\tau=0$, since we intend to study the correspondence with the
boundary state and therefore treat a closed string emitted by the
D-brane. Expand the action (\ref{Ssigma}) around a
classical configuration $\bar{X}$ as
\begin{eqnarray}
i\pi S[\bar{X}+\xi]
=&&\hs{-5mm}
  \left[\;\;
    \Half \int \! d^2\sigma\; \p_\alpha \bar{X}_M \p^\alpha \bar{X}^M 
    -i \oint \! d\sigma\;
    (\bar{X}^\mu)' A_\mu [\bar{X}]
  \;\;\right]    
\nn\\
&& \hs{-5mm}
+
\left[
\;\;\int\! d^2\sigma \;
\p_\alpha \bar{X}_M \p^\alpha \xi^M 
-i \oint\! d\sigma\;
\xi^\mu(\bar{X}^\nu)' F_{\mu\nu}[\bar{X}]
\;\;\right]
\nn\\
\label{expand}
&&\hs{-5mm}
+
\left[\;\;
\Half\int\! d^2\sigma \;
\p_\alpha \xi_M \p^\alpha \xi^M
-\frac{i}{2}
\oint\! d\sigma\;
\;
 \xi^\mu(\xi^\nu)' F_{\mu\nu}[\bar{X}]
\;\;\right]
\nn\\
&&\hs{-5mm}
-i
\oint\! d\sigma
\left(\Half 
\xi^\rho \xi^\nu (\bar{X}^\mu)' \p_\nu
F_{\rho\mu}[\bar{X}]
\right.
\nn\\
&&
\hs{10mm}
+\left.
{1\over 3} \xi^\nu \xi^\rho (\xi^\mu)' \p_\nu F_{\rho\mu}[\bar{X}]
+
{1\over 8}
\xi^\delta\xi^\rho\xi^\nu(\xi^\mu)' 
\p_\delta\p_\rho F_{\nu\mu}[\bar{X}]
+\cdots
\right),
\end{eqnarray}
where the fluctuation around the classical configuration is denoted as
$\xi$ and we have adopted Euclidean formalism, $i\tau \rightarrow
\tau$.  
In this expansion, we are working in slowly varying fields, {\it i.e.}
derivative expansion. 

In the above equation (\ref{expand}), terms in the second line
vanishes due to the equations of motion for the classical
configuration $\bar{X}$.  
The terms in the third line gives a kinetic term (propagator) for the 
scalar field $\xi$. This propagator depends on the constant field
strength  $F$ on the boundary. Usually in the string $\sigma$ model
calculation \cite{dorn,ACNY}, one evaluates a vacuum graph using the
vertex in the fourth 
line by the contraction
\begin{eqnarray}
  \label{Wick}
\wick{2}{
\oint \!d\sigma\;
<*\xi^\rho >*\xi^\nu (\bar{X}^\mu)' \p_\nu
F_{\rho\mu}[\bar{X}]
}.
\end{eqnarray}
This quantity is in proportion to $(\bar{X})'$, therefore the
divergence of this one-loop vacuum graph is interpreted as the
one-loop divergent quantum contribution to the gauge field $A_\mu[X]$ 
in the coupling $\oint\! d\sigma\; (\bar{X}^\mu)' A_\mu [\bar{X}]$
which appears in the first line in eq.\ (\ref{expand}). 

\begin{figure}[htdp]
\begin{center}
\begin{minipage}{80mm}
\begin{center}
\leavevmode
\epsfxsize=60mm
\put(70,-10){$\p\p F$}
\epsfbox{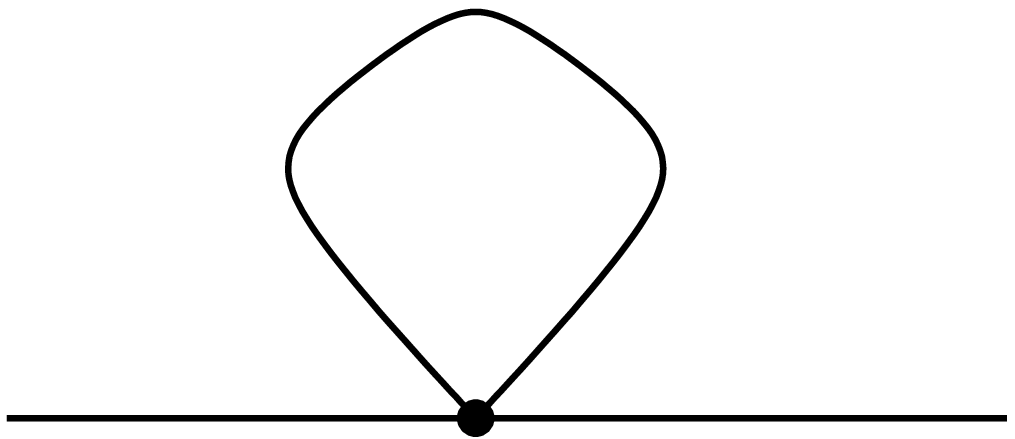}
\caption{Contribution of the $\p\p F$ term to the propagator with
  one-loop. The solid lines are propagation of the 
  fluctuation $\xi$, and the interaction vertex $\p\p F$ is on the
  boundary.} 
\label{loop1}
\end{center}
\end{minipage}
\hspace{5mm}
\begin{minipage}{70mm}
\begin{center}
\leavevmode
\epsfxsize=70mm
\put(118,40){$\p F$}
\put(60,40){$\p F$}
\epsfbox{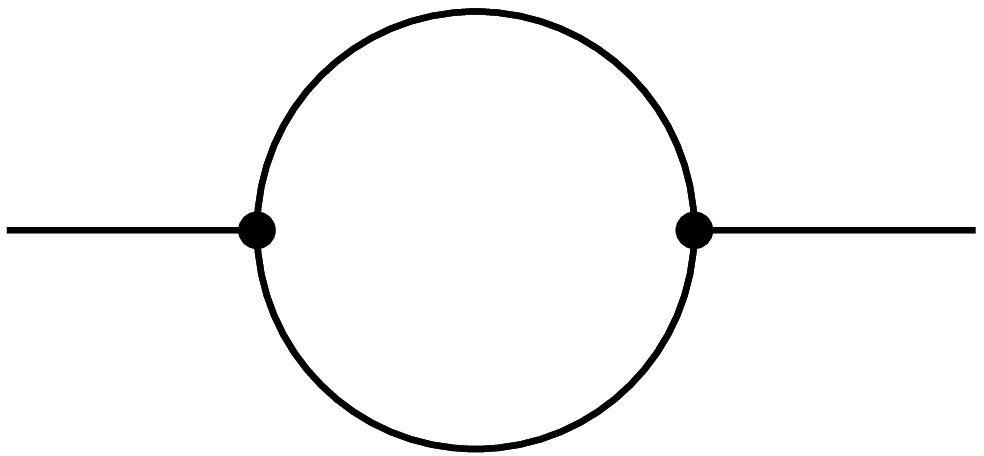}
\caption{Contribution of the $(\p F)^2$ term to the propagator with
  one-loop.} 
\label{loop2}
\end{center}
\end{minipage}
\end{center}
\end{figure}

However, when we ignore the first line and only consider the dynamics
of the scalar field $\xi(\sigma,\tau)$, it is natural to calculate the
renormalization of the propagator, three point vertex, and so on.
Let us concentrate on the one-loop correction to the
propagator. Using the couplings in the fifth line in
eq.\ (\ref{expand}), at one-loop level we have two relevant Feynman
graphs shown in Fig.\ \ref{loop1} and \ref{loop2},
which bring out divergences in the sector of the propagator. Though
generally the divergence in the two-point function is absorbed into
the renormalization factor of 
the wave function, now according to the philosophy of the string
$\sigma$ model, we expect that these divergences can be absorbed into
the redefinition of the gauge field strength in the propagator, with
the form (\ref{Fren}). 

We demonstrate here the calculation of the $\p\p F$ divergences
depicted in Fig.\ \ref{loop1}. With use of the 
oscillator representation of the boundary condition for the constant
field strength (\ref{(P-FX)B}),
\begin{eqnarray}
  \alpha_{n\mu}^{(-)}=-\calO_\mu^{\;\;\nu}\alpha_{-n\nu}^{(+)},
\end{eqnarray}
we obtain an explicit form of $X$ on the boundary as 
\begin{eqnarray}
X_\mu = x_\mu + \frac{i}{2}\sum_{n\neq 0}\alpha^{(+)}_{n\nu}
\left(
  z^n\delta_\mu^{\;\;\nu} + \bar{z}^{-n}\calO_\mu^{\;\;\nu}
\right)
\end{eqnarray}
where $z\equiv \exp(\tau+i\sigma)$, and $\tau$ is put to zero (the
boundary value) later. Then the correlation function is given by
\begin{eqnarray}
\lefteqn{  \bra{0}{\rm T}\left(
X_\mu(z,\bar{z})X_\nu(w,\bar{w})
\right)\ket{0}
}\nn\\
&& \hspace{-5mm}=-\frac14
\left[
  \eta_{\mu\nu}
  \left\{
    \log\left(1\!-\!\frac{\bar{w}}{\bar{z}}\right)
    + \log\left(1\!-\!\frac{z}{w}\right)
  \right\}
  + \calO_{\mu\nu}\log\left(1\!-\!\frac{1}{\bar{z}w}\right)
  + \calO_{\mu\nu}\log\left(1\!-\!\bar{w}z\right) 
\right].
\end{eqnarray}
This should be evaluated at the boundary $\tau_z, \tau_w \rightarrow
0$. Extracting the short distance behavior 
($\delta\equiv\sigma_z - \sigma_w$), we have the boundary correlators
as
\begin{eqnarray}
  \langle \xi_\mu(z)\xi_\nu(w) \rangle_{\rm boundary}
=C_{\mu\nu}\log \delta + {\rm finite},\\
  \langle \xi_\mu(z) \p_\sigma \xi_\nu(w) \rangle_{\rm boundary}
=C_{\mu\nu}\frac{-1}{\delta} + {\rm finite}, \label{contxx}\\
{\rm where} \qquad C_{\mu\nu}\equiv -\Half
\left(
  \eta_{\mu\nu} + \Half\calO_{\mu\nu}+\Half\calO_{\nu\mu}
\right).
\end{eqnarray}

There are 
${\scriptsize \left(\begin{array}{c}\!4\!\\\!2\!\end{array}\right)}
=6$ ways to contract two operators in the vertex 
${1\over 8}\oint \xi^\delta\xi^\rho\xi^\nu(\xi^\mu)' 
\p_\delta\p_\rho F_{\nu\mu}[\bar{X}]$ appearing in fig.\ \ref{loop1}. 
Noting the correspondence
\begin{eqnarray}
\log \delta : &&\hspace{-7mm}\quad
\lim_{\delta\rightarrow 0}  \log(1-e^{i\delta}) = -
\lim_{\delta\rightarrow 0}
\left(
  e^{i\delta} + \Half e^{2i\delta} + \frac13 e^{3i\delta} + \cdots
\right)
=-\sum_{n\geq 1}\frac1n = -\zeta(1),\\
\frac{1}{\delta} : &&\hspace{-7mm}\quad
\lim_{\delta\rightarrow 0}  \frac{-ie^{i\delta}}{1-e^{i\delta}} 
= -i
\lim_{\delta\rightarrow 0}
\left(
  e^{i\delta} + e^{2i\delta} + e^{3i\delta} + \cdots
\right)
=-i\sum_{n\leq 1}1 = -i\zeta(0),
\end{eqnarray}
one can see that the contraction (\ref{contxx}) brings about no
divergence after the $\zeta$ function regularization. Therefore we
need only to pick up half (three) of the contractions as
\begin{eqnarray}
  \frac{-i}{8}\oint\! d\sigma \; \p_\delta\p_\rho F_{\nu\mu}
\biggl(
\wick{2}{<*\xi^\delta>*\xi^\rho} \xi^\nu(\xi^\mu)'
+ 
\xi^\delta\wick{2}{<*\xi^\rho>*\xi^\nu}(\xi^\mu)'
+ 
\wick{2}{<*\xi^\delta \xi^\rho>*\xi^\nu}(\xi^\mu)'
\biggr)
\nn\\
=\frac{i}{4}\log\delta \oint\! d\sigma \; 
\xi^\mu(\xi^\nu)'
\left(
  \frac{1}{1\!\!+\!\!F}
\right)^{\rho\delta}\p_\delta\p_\rho F_{\mu\nu}.
\end{eqnarray}
Comparing this result with the propagator in eq.\ (\ref{expand}), 
the one-loop divergence in the propagator is absorbed into the
redefinition of the field strength as 
\begin{eqnarray}
F^{\rm red}_{\mu\nu}(x) \equiv F_{\mu\nu}(x)+ {1\over 2}\zeta(1)
\left( {1\over 1\!\!+\!\!F}\right)^{\rho\delta}
\p_\rho\p_\delta F_{\mu\nu}.
\end{eqnarray}
This agrees with eq.\ (\ref{Fren}), hence the divergences in the
boundary state correspond to the quantum loop divergence in the
propagator in the string $\sigma$ model.

The exponent in the boundary state (\ref{bou}) consists of the 
quadratic combination of the oscillators, therefore the boundary state 
respects the properties of the propagator in the string $\sigma$
model. Here we start from the single $\sigma$ model action
(\ref{Ssigma}) and expand it, hence it is plausible that the
divergences of both the propagator and the vacuum graph can be
absorbed into the same redefinition of the gauge field.

\section{Corrections to the D-brane action}
\label{sec:correction}

Having defined the boundary state $\bnewA$ for an arbitrary
configuration of the gauge field $A_\mu(x)$, 
our next task in this section is to extract a particular part of the
boundary state which describes the interaction with the massless modes 
of the closed string. This contributes to the corrections to the
D-brane action.

As confirmed in ref.\ \cite{our}, the couplings between the 
gauge field on the D-brane and the closed string field are described
in terms of SFT as $\braket{\Phi}{\BF}$. Thus this interaction (source
term) is naturally generalized to the one with non-constant field
strength:
\begin{eqnarray}
\label{braket}
\braket{\Phi}{\Bnew}= 
 \braket{\Phi}{B(F^{\rm red}(x))} 
+ \bra{\Phi}\left[
\begin{array}{l}
\mbox{creation operators}\\
\mbox{with finite coefficients}
\end{array}
\right]
\bFx.
\end{eqnarray}
Here we have substituted the representation (\ref{Bcreate}) of 
$\bnew$, therefore now we adopt the approximation
considered in the previous section.

The first term on the RHS of eq.\ (\ref{braket}) gives the same
contribution to the part of the D-brane action linear in the closed
string massless fields, as the one obtained in the case of the
constant field strength $\Phi\cdot\BF$ in ref.\ \cite{our}.
This is because, after putting the coefficient of the $\zeta(1)$ term
in eq.\ (\ref{Aren}) to zero, then we see $F^{\rm red}=F$. The result is
\cite{our} 
\begin{eqnarray}
\lefteqn{
\braket{\Phi}{B(F^{\rm red}(x))} 
= {-T_p\over 2}\int\!d^{p+1}x\;
\biggl(-\det \Bigl(\eta + F(x)\Bigr)\biggr)^{-\zeta(0)}
}\hspace{35mm}\nonumber\\
&&\biggl[
\tr \Bigl(h\!+\!b\Bigr)\Bigl(1\!+\!F(x)\Bigr)^{-1}\! + 
\Bigl\{
-2+{4\over d\!-\!2}\tr \Bigl(1\!+\!F(x)\Bigr)^{-1}
\Bigr\}D
\biggr]. 
\label{result}
\end{eqnarray}
Here $h_{\mu\nu}$ is the fluctuation of the metric field, $b_{\mu\nu}$
is the Kalb-Ramond two-form field, and $D$ is the dilaton field.
The divergent $\zeta(0)$ is regularized by the zeta-function
regularization. The resultant quantity (\ref{result}) is a part of 
the well known D-brane action linear in the closed 
string massless fields,\footnote{
The terms of higher powers in the closed string fields $h_{\mu\nu}$,
$b_{\mu\nu}$ and $\phi$ which exist in the ordinary D-brane action
are expected to be reproduced by performing the integrations over the
closed string massive modes.
Note that thus we obtain a D-brane action of a form {\it before}
integrating out the massive modes, {\it i.e.} we can read couplings
between the open string gauge field strength $F$ and closed string
massive modes.}
\begin{eqnarray}
-T_p\int\! d^{p+1}x 
\; e^{-D}\sqrt{-\det \Bigl(G_{\mu\nu} +b_{\mu\nu} 
+F^{\rm red}_{\mu\nu}(x)\Bigr)}
\Biggm|_{{\rm linear \; in}\; D,h_{\mu\nu},b_{\mu\nu}}.
\label{DBI}
\end{eqnarray}
The metric in this action is the string metric,\footnote{
The metric field in the D-brane action is defined by an induced one 
on the D-brane, and we have investigated its dependence in ref.\
\cite{our} by considering the tilt of the D-brane. We have seen the
perfect agreement with the string $\sigma$ model approach, in the case 
of the constant field strength and the constant tilt angle. 
Further generalization on general deformation of the D-brane surface
will be presented in our next paper \cite{KH}.}
which is related to the fluctuation $h$ of the Einstein metric
through Weyl rescaling as 
\begin{eqnarray}
G_{\mu\nu} = e^{4D/(d-2)}(\eta_{\mu\nu}+h_{\mu\nu}).
\end{eqnarray}

The second term in eq.\ (\ref{braket}) gives {\it finite} corrections
to the D-brane action, that include two derivatives acting on the
field strength $F(x)$. In order for the bracket to be expressed in
terms of, for example, the massless component fields of the closed
string, the oscillator contraction should be performed. 
Explicitly, the second term on the RHS of eq.\ (\ref{Bcreate}) reads
\begin{eqnarray}
  \label{finite}
\lefteqn{
\left[
\begin{array}{l}
\mbox{creation operators}\\
\mbox{with finite coefficients}
\end{array}
\right]
\bF
}\nn\\
&&
=-{1\over 8}\;
\p_\rho\p_\delta F_{\mu\nu}\;
\left[
J^{\kappa\delta}J^{\rho\lambda}J^{\mu\nu}
\right]
\alpha^{(-)}_{-1\;\kappa}
\alpha^{(+)}_{-1\;\lambda}\bF
\nn\\
&&\;\;\;\;
+\; {1\over 64}\;
\p_\gamma F_{\alpha\beta}\p_\rho F_{\mu\nu}\;
\left[
J^{\rho\gamma}J^{\beta\nu}(J^{\mu\alpha}+J^{\alpha\mu})
\right]\bF\nn\\
&&\;\;\;\;
+\; {1\over 32}\;
\p_\gamma F_{\alpha\beta}\p_\rho F_{\mu\nu}\;
\left[
2J^{\kappa\mu}J^{\nu\beta}J^{\gamma\rho}J^{\alpha\lambda}
-J^{\alpha\mu}J^{\nu\beta}J^{\kappa\rho}J^{\gamma\lambda}
\right]
\alpha^{(-)}_{-1\;\kappa}\alpha^{(+)}_{-1\;\lambda}\bF\nn\\[2mm]
&&\;\;\;\;+\;\cdots
\end{eqnarray}
where the matrix $J^{\mu\nu}$ is defined as 
\begin{eqnarray}
  \label{J}
J\;=\;1+\calO(x) \;=\; {2\over 1+F(x)}.
\end{eqnarray}
In (\ref{finite}) we have written explicitly only those terms which
contain the closed string massless modes, since these terms are
necessary for the evaluation of the D-brane action.
The omitted part ``$\cdots$'' in eq.\ (\ref{finite}) consists of an
infinite number of terms of massive excitations, 
such as $\sum_{l\geq 2}\alpha^{(-)}_{-l}\alpha^{(+)}_{-l} \ket{0}$ or
$(\alpha^\dagger)^n\ket{0}$ with $n\ge 3$. 

Note that the boundary state for the
constant field strength $\BF$ and the massless part of the string
field state $\Phi_{\rm massless}$ have even number of $\alpha$
oscillators. (We ignore the auxiliary fields (denoted as $b_M(x)$ and
$e_M(x)$ in ref.\ \cite{our}) in 
$\Phi$, since its bracket in eq.\ (\ref{braket}) vanishes due to the
contraction of the ghost oscillators.) Thus contribution of the terms
consisting of odd number of $\alpha$ oscillators disappears. With the
substitution of eq.\ (\ref{finite}), we find that the second term in
eq.\ (\ref{braket}), which is the correction to the D-brane action,
equals 
\begin{eqnarray}
\lefteqn{
{T_p\over 64}
\sqrt{-\det (\eta+F)}
\Bigl(
h_{\lambda\kappa} +b_{\lambda\kappa}+{4\over d-2}\eta_{\lambda\kappa}D
\Bigr)
}\nn\\
&&
\times
\left[
\p_\rho \p_\delta F_{\mu\nu}
\left(
J^{\kappa\delta}J^{\rho\lambda}J^{\mu\nu}
\right)
-{1\over 4}
\p_\gamma F_{\alpha\beta}\p_\rho F_{\mu\nu}
\left(
2J^{\kappa\mu}J^{\nu\beta}J^{\gamma\rho}J^{\alpha\lambda}
-J^{\alpha\mu}J^{\nu\beta}J^{\kappa\rho}J^{\gamma\lambda}
\right)
\right]
\nn\\
&&
\hs{-10mm}
-
{T_p\over 256}
\sqrt{-\det (\eta+F)}
\left[
\left(
h_{\lambda\kappa} +b_{\lambda\kappa}
\right)J^{\kappa\lambda}
+\left(-4+
{4\over d-2}J_\lambda^{\;\;\lambda}
\right)
D
\right]
\nn\\
&&
\hs{60mm}\times
\p_\gamma F_{\alpha\beta}\p_\rho F_{\mu\nu}
\left[
J^{\rho\gamma}J^{\beta\nu}(J^{\mu\alpha}+J^{\alpha\mu})
\right].
\label{whresult}
\end{eqnarray}
The factor $\zeta(0)$ has been regularized in the same way as before.


\section{Conclusion and discussion}
\label{conclusion}

We have constructed a new boundary state which has the degrees of
freedom of an arbitrary configuration of the gauge field $A_\mu(x)$ on 
the D-brane. 
This state is defined by using the closed SFT gauge transformation
acting on 
a conventional boundary state, inspired from the fact that a gauge 
transformation of a string $\sigma$ model generates all modes of the 
gauge field. The new boundary state is BRST invariant and surely
satisfies the nonlinear boundary conditions with the non-constant
filed strength.  

The new boundary state includes divergences of the order $\zeta(1)$
and $\zeta(0)$. The $\zeta(0)$ term can be regularized by the
ordinary zeta-function regularization method, whilst 
the intrinsic divergence of $\zeta(1)$ can be absorbed into the
redefinition of the gauge field. This redefinition exactly coincides
with the one-loop calculation for the gauge field coupling in a
string $\sigma$ model. In order for the generalized boundary
state to be well-defined, the divergent quantity should vanish, hence
this constraint forces the gauge field to satisfy a differential
equation. This equation is identical with the conformal invariance
condition  $\beta(A)=0$ derived in the string $\sigma$ model
approach. The origin of this coincidence is studied in the $\sigma$
model side, and we have found that the divergences in the new boundary
state can be identified with the one-loop contribution to the
worldsheet propagator in the string $\sigma$ model.

Utilizing this boundary state, we have computed the derivative
correction terms in the D-brane action, by extracting the part
corresponding to the 
massless excitations of the closed string. After putting the
divergence in the redefined gauge field to zero, 
we have obtained couplings  
between the derivatives of the field strength ($\p\p F$ and $(\p
F)^2$) and the closed string massless fields ($h_{\mu\nu}$,
$b_{\mu\nu}$ and $D$).

It might be possible to apply the procedure given in ref.\ \cite{our}
to obtain a gauge invariant complete SFT action. Although this seems
to be straightforward, we find it actually complicated. Only with a
naive extension of the method in ref.\ \cite{our}, there may not be 
corrected Born-Infeld action whose variation under the SFT gauge
transformation cancels the surface term stemming from the source term
$\braket{\Phi}{\Bnew}$. 

There still remains many subjects to be studied. In this paper we have 
argued the non-constant mode of the field strength on the D-brane, but 
it is more intrinsic to introduce massive modes of the {\it open}
strings. Though the couplings to the massive modes of the closed
strings has been manifested in ref.\ \cite{our}, the boundary state
contains infinite number of product of the gauge field, which is a
sign of the massive open strings. To construct more general state, the
SFT gauge transformation will play an important role. The closed SFT
adopted in sec.\ \ref{sec:extending} has been generalized to a
open-closed mixed system \cite{Asa}, hence some correspondence between 
our source term and open-closed coupling in ref.\ \cite{Asa} may help
to understand the boundary state further. At least, the corrections
(\ref{whresult}) may be reproduced in the low energy sector of the
system of ref.\ \cite{Asa}.

One of the other subjects is concerning the supersymmetry. 
Unfortunately, in the SFT scheme it is hard to discuss the
supersymmetry because of some schematic problems \cite{susy}. A lot of
interesting aspects of D-branes are on the basis of the supersymmetry. 
We expect that the definition of the new boundary state given in this
paper can be generalized to the supersymmetric  one. This sort of
generalization is seen in ref.\ \cite{CLNY2}, in which the case of
constant field strength was treated. With use of the generalized
supersymmetric boundary state, one can calculate corrections to the
D-brane actions in the superstring case as well as various interesting 
quantities \cite{KH}.

\vspace{.7cm}
\noindent
{\Large\bf Acknowledgments}\\[.2cm]
I am indebted to H.\ Hata for his collaboration at the early stage of
this work. I would like to thank also A.\ Hashimoto, Y.\ Matsuo, and
M.\ Akashi for useful comments. I am grateful to K.\ Furuuchi and N.\
Ishibashi for valuable discussions and careful reading of this
manuscript.  This work is supported in part by Grant-in-Aid for
Scientific Research from Ministry of Education, Science, Sports and
Culture (\#3160). I appreciate hospitality of the organizers of Summer
Institute `99 where a part of this work was argued. 


\vs{3cm}

\noindent
{\Large\bf Appendix}

\vs{-5mm}

\appendix

\section{Calculation of the $\star$ product for an arbitrary %
  $\zeta_\mu(x)$} 
\label{app:star}

In order to view the meaning of
the unitary rotational operator $U[A]$,
it is important to check that the definition (\ref{Bnew}) really
follows from a gauge rotation ($\star$ product)
in SFT. What should be checked is 
\begin{eqnarray}
\bnew = \ket{B(F\!=\!0)} 
-2 \ket{\Lambda\Bigl( -A_\mu (x)\Bigr) \star B(F\!=\!0)}
\end{eqnarray}
for an infinitesimal $A_\mu (x)$ (as a parameter in the $\Lambda$), 
instead of the previous $\zeta_\mu(x)$. It is needed only to trace 
the app.\ B of ref.\ \cite{our}, for $\zeta_\mu (x)$ with arbitrary 
powers in $x$. We will present only the summary of the calculation. In 
this case we should take into consideration the $F_{p^2}$ term (eq.\
(B.6) in \cite{our}) 
\begin{eqnarray}
\exp {F_{p^2}} = \exp 
\left(-{1\over 4} \eta^{\mu\nu}p_\mu p_\nu 
\log \Biggl| {e\alpha_2 \over \epsilon} \Biggr|
+\cdots \right)                                          \label{Fp}
\end{eqnarray}
in the three string vertex operator $V_{123}$. This $F_{p^2}$ factor
is quadratic in $p$, thus when treating $\zeta_\mu$ linear in $x$ 
($ = i{\p \over \p p}$) as in \cite{our} its contribution disappears.
But for higher powers in $x$, the logarithmic divergence $\log 
\left|{\epsilon\over \alpha}\right|$ comes out. This divergences are 
absorbed into the normal ordering divergences of the $X$'s in 
$\zeta_\mu(X)$, but its finite part remains ambiguous. For example,
the gauge transformation parameter
\begin{eqnarray}
\zeta_\mu (x)\, = \,
\zeta_{\mu\nu\rho} x^\nu x^\rho +
\zeta_{\mu\nu\rho\delta} x^\nu x^\rho x^\delta
                                                   \label{zeta23}
\end{eqnarray}
results in the identity
\begin{eqnarray}
-2 \ket{\Lambda \star \BF} = 
\biggl[
{i\over\pi}\oint\!\p_\sigma X^\mu \zeta_\mu (X)
+c\cdot{i\over\pi}\oint\!\p_\sigma X^\mu 
(\zeta_{\mu\nu\rho}{}^\rho+\zeta_{\mu\rho\nu}{}^\rho
+\zeta_{\mu\rho}{}^\rho{}_\nu) X^\nu 
\biggr] \bF,
\label{c}
\end{eqnarray}
where $c$ is a {\it finite} unknown constant (contractions of 
$\zeta_{\mu\nu\rho\delta}$ comes from the flat metric $\eta^{\mu\nu}$
in eq.\ (\ref{Fp})).

In this paper we assume that $c=0$, since if we define 
$\bnew$ with non-zero $c$ using the above eq.\ (\ref{c}), the 
generalized nonlinear boundary condition (\ref{eigen}) does not
follow.  
After all, together with the above results and the definitions
(\ref{transnew}) (\ref{Bnew}), the gauge transformation of the
dynamical variable $A_\mu(x)$ of the new boundary state $\bnew$ 
generated by the SFT $\star$ product is (\ref{gaugetr}), for an
arbitrary number of powers in $x_\mu$.


\section{Calculation of the divergence in the boundary state}
\label{app:div}

In this appendix we present the calculation to obtain eq.\
(\ref{Bcreate}), changing all the oscillators into creation operators
using the boundary condition (\ref{bc}). 
We show here the calculation of only the last term in eq.\ (\ref{V}),
since this term is the most bothersome and other terms can be
evaluated in the same way. Considering the corresponding
mode of the gauge field 
\begin{eqnarray}
  A_\mu(x) = \zeta_{\mu\nu\rho} x^\nu x^\rho, \qquad
{\rm with}\;\; \zeta_{\mu\nu\rho} = \zeta_{\mu\rho\nu},
\end{eqnarray}
the last term in eq.\ (\ref{V}) is written as 
\begin{eqnarray}
T\equiv \frac{-1}{32} \zeta_{\mu\nu\rho}\zeta_{\alpha\beta\gamma}
 \left[
   \sum_{n,m}\frac{1}{nm}
   \left(\alpha_{-n-m}^{(+)\mu}-\alpha_{n+m}^{(-)\mu}   \right)
   \left(\alpha_{n}^{(+)\nu}-\alpha_{-n}^{(-)\nu}   \right)
   \left(\alpha_{m}^{(+)\rho}-\alpha_{-m}^{(-)\rho}   \right)
 \right]
\hs{20mm}\nn\\
\times
 \left[
   \sum_{k,l}\frac{1}{kl}
   \left(\alpha_{-k-l}^{(+)\alpha}-\alpha_{k+l}^{(-)\alpha}   \right)
   \left(\alpha_{k}^{(+)\beta}-\alpha_{-k}^{(-)\beta}   \right)
   \left(\alpha_{l}^{(+)\gamma}-\alpha_{-l}^{(-)\gamma}   \right)
 \right].
\end{eqnarray}
This is evaluated on the boundary state $\bFx$. After changing all the
oscillators into creation operators, then we are led to the following
form symbolically:
\begin{eqnarray}
  T\bFx= 
\left[
(\alpha^\dagger)^6 + (\alpha^\dagger)^4
 + (\alpha^\dagger)^2 + c
\right]
\bFx.
\label{a6}
\end{eqnarray}
On the RHS, the first two terms have finite coefficients and
furthermore are not relevant for the massless modes in the closed
string excitations, hence it is not necessary to evaluate them
explicitly. The third term quadratic in $\alpha$ and the last constant
term $c$ contain divergences and also finite contributions.
The $(\alpha^\dagger)^2$ term is given as 
\begin{eqnarray}
(\alpha^\dagger)^2 =
& \displaystyle\sum_{m\!+\!n>0, m<0}
\alpha^{(-)\sigma}_{-n-m}\alpha^{(+)\delta}_{-n-m}
\Biggm[
&
\frac{2}{m(n\!+\!m)}
         (-J^{\sigma\mu}J^{\nu\alpha}J^{\beta\rho}J^{\gamma\delta}
          -J^{\mu\delta}J^{\alpha\nu}J^{\rho\beta}J^{\sigma\gamma})
\nn\\
& & 
+\frac{2}{n(n\!+\!m)}
         (-J^{\sigma\mu}J^{\nu\beta}J^{\alpha\rho}J^{\gamma\delta}
          -J^{\mu\delta}J^{\beta\nu}J^{\rho\alpha}J^{\sigma\gamma})
\nn\\
& & 
\;\;+\frac{-2}{mn}
         (-J^{\sigma\mu}J^{\nu\beta}J^{\gamma\rho}J^{\alpha\delta}
          -J^{\mu\delta}J^{\beta\nu}J^{\rho\gamma}J^{\sigma\alpha})
\Biggm]\nn\\
& +\displaystyle\sum_{m>0, n>0}
\alpha^{(-)\sigma}_{-n-m}\alpha^{(+)\delta}_{-n-m}
\Biggm[
&
\frac{-2}{m(n\!+\!m)}
         (-J^{\sigma\mu}J^{\nu\alpha}J^{\rho\beta}J^{\gamma\delta}
          -J^{\mu\delta}J^{\alpha\nu}J^{\beta\rho}J^{\sigma\gamma})
\nn\\
& & 
+\frac{-2}{n(n\!+\!m)}
         (-J^{\sigma\mu}J^{\nu\beta}J^{\rho\alpha}J^{\gamma\delta}
          -J^{\mu\delta}J^{\beta\nu}J^{\alpha\rho}J^{\sigma\gamma})
\nn\\
& & 
\;\;+\frac{2}{mn}
         (-J^{\sigma\mu}J^{\nu\beta}J^{\rho\gamma}J^{\alpha\delta}
          -J^{\mu\delta}J^{\beta\nu}J^{\gamma\rho}J^{\sigma\alpha})
\Biggm]
\nn\\
&+\;\;\mbox{other summations.}&
\label{summ}
\end{eqnarray}
Here ``other summations'' denotes the summation in the other four
regions in the $(m,n)$-plane divided by three lines: $m\!=\!0, n\!=\!0,
m\!+\!n\!=\!0$. The other summation is performed in the same manner
with slight changes of the indices. From this expression we can
extract the divergences $\zeta(1)$. For example, from the first
summation we have 
\begin{eqnarray}
  \zeta(1) \displaystyle\sum_{\wt{n}>0}\frac{2}{\wt{n}}
\alpha^{(-)\sigma}_{-\wt{n}}\alpha^{(+)\delta}_{-\wt{n}}
\Biggm[
-          (-J^{\sigma\mu}J^{\nu\alpha}J^{\beta\rho}J^{\gamma\delta}
            -J^{\mu\delta}J^{\alpha\nu}J^{\rho\beta}J^{\sigma\gamma})
\hs{10mm}\nn\\
+
           (-J^{\sigma\mu}J^{\nu\beta}J^{\alpha\rho}J^{\gamma\delta}
            -J^{\mu\delta}J^{\beta\nu}J^{\rho\alpha}J^{\sigma\gamma})
\Biggm],
\end{eqnarray}
where we have defined $\wt{n}\equiv n+m$. This is because, in this
case we should sum over $\wt{n}$ as the index of the oscillators, and
for a fixed $\wt{n}$, divergent quantities proportional to $\zeta(1)$
stem from the first and the second terms in the first summation in
eq.\ (\ref{summ}) (the factor of the third term $-2/nm$ does not give the
divergence). In the same way, one finds that the second summation in
eq.\ (\ref{summ}) does not bring about 
the divergence. Collecting all the divergent terms, we can verify the
following equality:
\begin{eqnarray}
  \mbox{divergent terms of}\;(\alpha^\dagger)^2 =
\ket{B_{\rm N}(F+\delta F)} - \ket{B_{\rm N}(F)}
\end{eqnarray}
with
\begin{eqnarray}
  \delta F^{\lambda\kappa}\equiv -\zeta(1)\cdot\Half
\zeta_{\mu\nu\rho}\zeta_{\alpha\beta\gamma}
\left[
J^{\alpha\mu}J^{\nu\beta}\eta^{\rho\lambda}\eta^{\gamma\kappa}
+J^{\mu\beta}J^{\gamma\nu}\eta^{\alpha\kappa}\eta^{\rho\lambda}
-J^{\nu\beta}J^{\gamma\mu}\eta^{\alpha\kappa}\eta^{\rho\lambda}
\right]-(\kappa\leftrightarrow\lambda).
\label{deltaF}
\end{eqnarray}
In the same way, the constant term $c$  in eq.\ (\ref{a6}) is
evaluated, and the result is  
\begin{eqnarray}
  \mbox{divergent terms of}\;c =
N(F+\delta F) - N(F)
\end{eqnarray}
with the above same $\delta F$ (\ref{deltaF}). 
These give the relation (\ref{Bcreate}), using the identification
\begin{eqnarray}
  \p_\gamma F_{\alpha\beta} = 
2(\zeta_{\alpha\beta\gamma}-\zeta_{\beta\alpha\gamma}).
\end{eqnarray}

\newcommand{\J}[4]{{\sl #1} {\bf #2} (#3) #4}
\newcommand{\andJ}[3]{{\bf #1} (#2) #3}
\newcommand{\AP}{Ann.\ Phys.\ (N.Y.)}
\newcommand{\MPL}{Mod.\ Phys.\ Lett.}
\newcommand{\NP}{Nucl.\ Phys.}
\newcommand{\PL}{Phys.\ Lett.}
\newcommand{\PR}{Phys.\ Rev.}
\newcommand{\PRL}{Phys.\ Rev.\ Lett.}
\newcommand{\PTP}{Prog.\ Theor.\ Phys.}
\newcommand{\HIKKO}{
H.\ Hata, K.\ Itoh, T.\ Kugo, H.\ Kunitomo and K.\ Ogawa}
\newcommand{\hep}[1]{{\tt hep-th/#1}}

\end{document}